\documentclass{aastex62}

\usepackage{amsmath}
\usepackage{txfonts}
\usepackage[T1]{fontenc}
\usepackage{url}
\usepackage{graphicx}
\usepackage[caption=false]{subfig}

\shortauthors{Brown et al.}
\begin{document}

\title{Oscillator Strengths for Ultraviolet Transitions in P {\small II} -- The Multiplet at 1308 \AA}

\correspondingauthor{R.B. Alkhayat} \\
\email {rabee.alkhayat@rockets.utoledo.edu} 

\author{M.S. Brown}
\affiliation{Department of Physics and Astronomy, University of Toledo, Toledo, OH 43606}
\author{R.B. Alkhayat}
\affiliation{Department of Physics and Astronomy, University of Toledo, Toledo, OH 43606}
\author{R.E. Irving}
\affiliation{Department of Physics and Astronomy, University of Toledo, Toledo, OH 43606}
\author{N. Heidarian}
\affiliation{Department of Physics and Astronomy, University of Toledo, Toledo, OH 43606}
\author{J. Bancroft Brown}
\affiliation{Department of Physics and Astronomy, University of Toledo, Toledo, OH 43606}
\author{S.R. Federman}
\affiliation{Department of Physics and Astronomy, University of Toledo, Toledo, OH 43606}
\author{S. Cheng}
\affiliation{Department of Physics and Astronomy, University of Toledo, Toledo, OH 43606}
\author{L.J. Curtis}
\affiliation{Department of Physics and Astronomy, University of Toledo, Toledo, OH 43606}

\begin{abstract}

We report lifetimes, branching fractions, and the resulting oscillator strengths for transitions within the P~{\small II} multiplet (3$s^2$3$p^2$ $^3P$ $-$ 3$s$3$p^3$ $^3P^{\rm o}$) at 1308 \AA. These comprehensive beam-foil measurements, which are the most precise set currently available experimentally, resolve discrepancies involving earlier experimental and theoretical results. Interstellar phosphorus abundances derived from $\lambda$1308 can now be interpreted with greater confidence.  In the course of our measurements, we also obtained an experimental lifetime for the 3$p$4$s$ $^3P_0^{\rm o}$ level of P~{\small IV}. This lifetime agrees well with the available theoretical calculation.
\end{abstract}


\section{Introduction} 

Singly$-$ionized phosphorus, which is the element's dominant charge state in neutral diffuse interstellar gas, is used to probe depletion onto interstellar grains in the Galaxy \citep*[e.g.,][]{jur78, duf86, jen86, leb05, car06, jen09} and in the Magellanic Clouds  \citep[e.g.,][]{mal01, tch15}. Knowledge of the amount of depletion in terms of gas density then allows studies of metallicity and the nucleosynthetic history in more distant galaxies and damped Lyman-$\alpha$ systems  \citep[e.g.,][]{mol01, lev02,pet02,des07,som15,dec16} via P~{\small II} absorption. Two transitions originating from the ground state ($\lambda\lambda$1153, 1302) in this ion are the most commonly used ones for Galactic studies \citep[e.g.,][]{jen86, car06, jen09}. These transitions have moderate to moderately large oscillator strengths ($f$-values), as noted in the most recent compilation of \citep{mor03}; the line at 1302 \AA\ is the weaker of the two. Morton recommended the theoretical $f$-values of \citep{hib88} for the multiplet ($\lambda$1308) containing the line at 1302 \AA. Hibbert's computed lifetimes for the upper levels 3$s$3$p^3$ $^3P_j^{\rm o}$ are in close agreement with the experimental results of \citet{liv75}, who quoted about 20\% accuracy. Other experimental results \citep{sav66,cur71,smi78} for this multiplet, however, differ considerably (factors of 2) from those of \citet{liv75}. The early theoretical work of \citet{bcs72} agreed reasonably well with the experimental results available at the time \citep{sav66,cur71}. The theoretical results of \citet{bra93} lie between those of \citet{cur71} and \citet{liv75}. More recent large-scale calculations do not provide a resolution.  While the results of \citet{tay03} and \citet{fro06} agree very well, the $f$-values are nearly twice those given by \citet{hib88}. The importance of P~{\small II} oscillator strengths in astronomical studies is discussed in the recent compilation by \citet{cas17}, who prefer using the results from \citet{fro06}.

These conflicting results have forced astronomers to derive empirical $f$-values from interstellar spectra, but the situation remains unsatisfactory. Based on their comparison of $b$-values for interstellar S~{\small  II} and P~{\small II} lines, \citet*{har86} suggested a value of 9.2 for the ratio of $f$-values for the lines at 1153 and 1302 \AA.  \citet{duf86}, on the other hand, adopted earlier calculations by \citet{hib86} and in the analysis of their data that resulted in a ratio of 22. This is similar to the ratio obtained by \citet{hib88}. For comparison, \citet{bra93} obtained a ratio of 11.2, \citet{tay03} a ratio of 11.5, and \citet{fro06} a ratio of 12.1. 

We seek a resolution to the question of the most appropriate $f$-values to use for interstellar studies by extending our earlier results on the multiplet containing the line at 1153 \AA\ \citep{fed07} through measurements of the pertinent data for the multiplet $\lambda$1308. Through beam-foil spectroscopic techniques, we present experimental lifetimes and branching fractions to derive $f$-values. Our empirical results are the most comprehensive to date and are of sufficient accuracy to discern the most consistent theoretical efforts.

\section{Experimental Details} 

The measurements were performed at the Toledo Heavy Ion Accelerator. Details describing this facility can be found in our earlier papers \citep[e.g.,][]{fed92,haa93, sch00}. Here we focus on the specfics relevant to the P~{\small II} data for the multiplet at $\lambda$1308.  As in our earlier work on P~{\small II} \citep{fed07}, phosphorus ions were produced utilizing a low-temperature oven and subsequent charging through interactions in an argon plasma.  Lifetime measurements and the spectra for branching fractions were obtained using energies of 170 and 220 keV. Because the lines in this multiplet were relatively weak, only forward decay curves were acquired (i.e., moving away from the slit of the monochromator). Beam divergence, foil thickening, and nuclear scattering were investigated while interpreting our data \citep[see][]{fed07}. Our ion source was typically set to produce  P$^+$ prefoil beam currents of 100 nA.  These ions passed through carbon foils whose thicknesses were 2.4 $\mu$g cm$^{-2}$ and emerged both in a variety of charge states and excited states.  An Acton 1 m normal-incidence vacuum ultraviolet monochromator with 2400 line mm$^{-1}$ gratings blazed at 800 \AA\ was used for the measurements. Lifetime measurements and the spectrum of P~{\small IV} were obtained with one grating; the P~{\small II} spectra were obtained with a newer grating of the same specifications. The only noticeable effect involved the instrumental linewidth, which changed from 0.22 to 0.27 \AA. A Galileo channeltron was utilized to detect the emitted radiation at the exit slit of the monochromator.

Blending among spectral features and the relative weakness of the lines of interest prevented us from performing a complete set of measurements. We instead focused on those which would yield accurate results. Lifetimes for the $J$ $=$ 1 and 2 levels in the upper state 3$s$3$p^3$ $^3P^{\rm o}$ were obtained from decay curves involving transitions at 1301.87 and 1310.70 \AA, respectively, to the ground state 3$s^2$3$p^2$ $^3P$. Systematic effects were studied by the acquisition of decay curves for $J$ $=$ 2 at two energies. Multiexponential fits were used to extract lifetimes. However, in most instances a single exponential sufficed. Furthermore, the error bars associated with the lifetimes take into account both statistical and experimental uncertainties. The decay curve for the $J$ $=$ 2 line obtained in the forward direction was best fit by two exponentials; we ascribe the short-lived decay component of the fit to a transition in P~{\small IV} at 655 \AA\ seen in second order (3$p^2$ $^3P_1$ -- 3$p$4$s$ $^3P_0^{\rm o}$).  We only saw the effects of P~{\small IV} in one decay curve because the feature is present on the red side of the P~{\small II} line and therefore depended on the placement of the grating when peaking up on $\lambda$1310.  Additional measurements at 205 keV were made on the P~{\small IV} line at 655 \AA\ for further confirmation of its short lifetime. Figure~\ref{fig:f1} shows the decay curve for the $^3P_2$ $-$ $^3P_2^{\rm o}$ line of P~{\small II} at 1310 \AA\ at a beam energy of 170 keV. 

Cascades from higher lying levels, which affect the populations of the levels of interest, could impact our lifetime measurements. Large-scale computations \citep{hib88, tay03, fro06}, however, suggest that repopulation is not a concern in our experiment. These calculations indicate that decays from the 3$s^2$3$p$4$p$ $^3D$, $^3P$, and $^3S$ states may repopulate the 3$s$3$p^3$ $^3P^{\rm o}$ state, with typical lifetimes of 7 to 12 ns. We do not find clear evidence for lifetimes in this range from our multiexponential fits at the precision of our measurements. More importantly, \citet{hib88} and \citet{fro06} determined multiplet Einstein $A$ coefficents for transitions between these states that were less than a few times $10^7$ s$^{-1}$; this suggests the transitions may be too weak to play a significant role in the repopulation of the levels in question.

When more than one decay channel is present, a combination of lifetimes and branching fractions are needed to derive oscillator strengths. The theoretical results of \citet{hib88}, \citet{tay03}, and \citet{fro06} indicate that intercombination lines arising from the 3$s$3$p^3$ $^3P^{\rm o}$ state are weak and are below the sensitivity of our experiment.  Therefore, we focused on branching among the dipole-allowed transitions between 1300 and 1312 \AA. Two scans were required to cover the multiplet in which there was an overlap in the region between 1304 and 1306 \AA. The overlap region was utilized to normalize the spectra covering the multiplet. The added signal in this portion of the spectrum aided in fitting the blend of lines.  A further complication was the numerous weak lines from the P~{\small IV} multiplet $\lambda$655 seen in second order at these wavelengths, as well the P~{\small I} lines appearing in first order at 1305.0, 1305.7 and 1306.0 \AA\ \citep{kel87}. In order to minimize effects of the rapidly decaying P~{\small IV} features, spectral scans were acquired 4 mm upstream from the monochromator entrance slit at a beam energy of 170 keV.

As a guide to help determine the amount of contamination from the P~{\small IV} lines in P~{\small II} spectra, a spectrum from 650 to 658 \AA\ with a beam energy of 205 keV was acquired with the foil position zero mm upstream of the monochromator entrance slit (see Figure~\ref{fig:f2}). The P~{\small IV} spectrum was fit by allowing the dispersion, line intensities, and the multiplet wavelength to vary while keeping the line separations fixed to the wavelengths in \citet{kel87}. The intensities from this fit were scaled to those seen in the P~{\small II} spectrum at 4 mm upstream with a dispersion of 0.27 \AA, and the scaled P~{\small IV} lines were subtracted from the P~{\small II} spectrum. The impact of the contamination from P~{\small I} lines on the P~{\small II} multiplet under study was deduced from a spectrum obtained at 120 keV at zero mm offset. The least blended P~{\small I} line at 1306.0 \AA\ was fit and the intensities for the other lines scaled to the intensities in \citet{kel87} before subtracting the P~{\small I} lines from the P~{\small II} spectrum at 4 mm. Figure 3a shows the uncorrected P~{\small II} spectrum with contaminating features and the corrected P~{\small II} spectrum is displayed in Figure 3b.

Branching fractions are usually determined by measuring the relative integrated intensities from lines with a common upper level through Gaussian fits. The fitting procedure indicated that the line widths were indistinguishable from one another; we, therefore, relied on the intensities. The spectral interval is small enough that systematic differences in instrumental response are not a concern. For determining the uncertainty in branching fractions, the statistical error estimated from the uncertainty in P~{\small II} line intensities was added in quadrature to the uncertainty in lifetime for the corresponding upper level. The systematic error from the P~{\small I} and P~{\small IV} blends was also considered; it was inferred from relative intensities of the contaminating lines to the P~{\small II} lines, which did not exceed 5 to 10\%. These systematic errors were added in quadrature to the statistical errors to determine the total errors in the branching fractions.

\section{Results and Discussion}

Tables~\ref{table:t1} through~\ref{table:t3} show the results of our lifetime measurements, branching fractions, and derived oscillator strengths, respectively. In Table~\ref{table:t1}, we present our lifetime measurements as well as comparisons with earlier experiments and theoretical calculations. The lifetime of the 3$s$3$p^3$ $^3P^{\rm o}$ $J_u$=2 level was obtained from the weighted average of measurements of the  $\lambda$1310 line at two beam energies, 170 and 220 keV. Figure~\ref{fig:f1} shows the decay curve for the line at 1310 \AA\ at a beam energy of 170 keV. Our lifetime measurements are most consistent with the beam-foil results of \citet{liv75} and with the theoretical calculations of \citet{bra93}, \citet{tay03}, and \citet{fro06}. These theoretical efforts were based on versions of the multiconfiguration-Hartree-Fock (MCHF) technique, with corrections for relativistic effects. However, the two experiments using the phase shift method \citep{sav66,smi78} yielded much shorter lifetimes. Our experimental values differ slightly from the lifetimes given by \citet{hib88}. We note that $\lambda$1304.49  associated with $J_u=0$ is too blended with $\lambda$1304.68, making it difficult to measure the lifetime for this level. Therefore, we do not list a multiplet value in our results for oscillator strength. However, the theoretical calculations \citep{hib88, tay03} predicted that the levels with higher angular momenta have longer lifetimes.

Our lifetime for the 3$p$4$s$ $^3P_0^{\rm o}$ level of P~{\small IV} is $0.27 \pm 0.04$ ns, a value that agrees with the theoretical prediction of \citet{gup02} (0.2205 ns), who used the CIV3 code of \citet{hib75}.  The isolated measurement confirms our supposition that the short-lived decay found for the $J$ $=$ 2 level of P~{\small II} 3$s$3$p^3$ $^3P^{\rm o}$ (see Figure~\ref{fig:f1}) arises from P~{\small IV}. The result for P~{\small IV} leads us to the following suggestions regarding earlier experimental determinations. First, the shorter lifetime obtained by \citet{cur71}, also based on beam-foil spectroscopy, could be the result of contamination from the P~{\small IV} decay.  Second, \citet{liv75} noted the presence of a second decay besides the one associated with P~{\small II} with a lifetime of $1.2 \pm 0.4$ ns that they ascribed to a blend with O~{\small I}.  Within the precision of their measurements, the shorter lifetime could instead arise from P~{\small IV}.

\citet{hib88}, \citet{tay03}, and \citet{fro06} provided a complete set of results for the multiplet at 1308 \AA, from which branching fractions can be inferred.  The branching fractions from the three calculations agree well and show slight differences from those obtained from LS coupling rules (see Table~\ref{table:t2}). LS coupling gives relative strengths of 0.750 and 0.250 for the transitions to $J_l$ $=$ 2 and $J_l$ $=$ 1.  \citet{hib88} found values of 0.758 and 0.242, while \citet{tay03} gives branching fractions of 0.756 and 0.241 and \citet{fro06} 0.756 and 0.244. Some minor variation is seen among the theoretical calculations for the $J_u$ $=$ 1 branching fractions, where LS coupling gives 0.417, 0.250, and 0.333 for decays to $J_l$ $=$ 2, 1, and 0. \citet{hib88} obtained respective values of 0.378, 0.276, and 0.346, while those of \citet{tay03} are 0.391, 0.270, and 0.339 and those of \citet{fro06} are 0.414, 0.254, and 0.332. Also, shown are the results of a semi-empirical calculation \citep{ban11} in which the mixing between LS-coupled basis states is determined from the observed energy levels. This simple approach has worked well for forbidden lines in a number of sequences but clearly fails to account precisely for the configuration interactions in the present case.

Although the three computations yielded very similar branching fractions, we obtained experimental values to check for the possibility of unexpected anomalies and they are also shown in Table~\ref{table:t2}.  The results from fitting our spectrum of the multiplet are consistent with these predictions, but they do not favor one over another because the experimental uncertainties are too large. The experimentally determined branching fractions for the $J_u$ $=$ 2 level are $0.22 \pm 0.02$ ($J_l$ $=$ 1) and $0.78 \pm 0.05$ ($J_l$ $=$ 2), while for the $J_u$ $=$ 1 level they are $0.36 \pm 0.03$ ($J_l$ $=$ 0), $0.27 \pm 0.03$ ($J_l$ $=$ 1), and $0.37 \pm 0.04$ ($J_l$ $=$ 2). We did not apply a correction to the intensities between spectra at 4 mm upstream and at zero offset relative to the entrance slit because our precision for lifetimes could not discern the very small variation with $J_u$ seen in the calculations.

Finally, we compare our results on oscillator strength derived from our lifetimes and branching fractions in Table~\ref{table:t3}. Our experimental results for oscillator strengths are in very good agreement with those of \citet{tay03} and \citet{fro06}. The calculations from \citet{hib88}, which formed the basis of Morton's (2003) compilation, are lower by about 30\%. These oscillator strengths are smaller because of the  cancellations in the dipole matrix elements due to the strong configuration interactions between 3$s$3$p^2$ and 3$s^2$3$pnd$. Moreover, the ratio of our experimental oscillator strengths for P~{\small II} $\lambda$1153 obtained from \citet{fed07} and P~{\small II} $\lambda$1302 acquired from this work is $13.9 \pm  2.0$, which is in very good agreement with the ratio given by MCHF calculations \citep{bra93, tay03, fro06}. Thus, our measurements provide support for the assessment on P~{\small II $f$-values given by \citet{cas17}.

\section{Conclusions}

Results of beam-foil measurements yielding lifetimes and branching fractions for the transitions associated with the P~{\small II } multiplet at 1308 \AA\ were presented.  These were combined to yield oscillator strengths that are needed for studies of the phosphorus abundance in the diffuse interstellar medium of our Galaxy and more distant ones. These experimental results represent the most precise values for this multiplet to date. Our branching fractions agree quite well with LS coupling rules but cannot differentiate between them and theoretical calculations given by \citet{hib88}, \citet{tay03}, and \citet{fro06}. Our results for lifetimes and oscillator strengths agree with the MCHF results of \citet{tay03} and \citet{fro06}. Furthermore, the $f$-value ratio from our beam foil techniques for lines at 1153 and 1302 \AA\ is $13.9\pm 2.0$, very similar to the ratio obtained from MCHF calculations. The consistency now revealed between experiment and theory for these transitions in P~{\small II } will provide more secure phosphorus abundances for interstellar matter. 

\acknowledgments
This work was supported by NASA grant NNG06GC70G. J. Bancroft Brown and R.E. Irving acknowledge support by the National Science Foundation under Grant No. PHY- 0648963. We thank Dave Ellis for useful conversations.


\begin{deluxetable}{cccccccccc} 
\tablecolumns{10}
\tablewidth{0pt}
\tabletypesize{\footnotesize}
\tablecaption{P~{\small II} Lifetimes for 3$s$3$p^3$ $^3P^{\rm o}$ Levels}
\tablehead{\colhead{$J_u$} & \multicolumn{9}{c}{$\tau$ (ns)} \\
\cline{2-10} \\
\colhead{} & \colhead{Present} & \colhead{SL\tablenotemark{a}} & \colhead{CMB\tablenotemark{b}} & \colhead{LKIP\tablenotemark{c}}&\colhead{S\tablenotemark{d}} & \colhead{H\tablenotemark{e}} & \colhead{BMF\tablenotemark{f}} & \colhead{T\tablenotemark{g}} & \colhead{FTI\tablenotemark{h}}
}
\startdata
1 & $14.0 \pm 0.8$ & $6.4 \pm 0.8$ & $9.0 \pm 0.5$ & $15 \pm 4$ & $5.4 \pm 0.6$ & 20.7 & $\ldots$ & 12.5 & 12.2 \\
2 & $14.6 \pm 0.5$ & $6.4 \pm 0.8$ & $9.0 \pm 0.5$ & $15 \pm 4$ & $5.4 \pm 0.6$ & 22.6 & 12.3 & 13.4 & 13.0 \\
\enddata
\tablenotetext{a}{\citet{sav66} $-$ phase shift experiment.}
\tablenotetext{b}{\citet{cur71} $-$ beam-foil experiment.}
\tablenotetext{c}{\citet{liv75} $-$ beam-foil experiment.}
\tablenotetext{d}{\citet{smi78} $-$ phase shift experiment.}
\tablenotetext{e}{\citet{hib88} $-$ configuration interaction calculation.}
\tablenotetext{f}{\citet{bra93} $-$ multi-configuration Hartree-Fock calculation.}
\tablenotetext{g}{\citet{tay03} $-$ multi-configuration Hartree-Fock calculation.}
\tablenotetext{h}{\citet{fro06} $-$ multi-configuration Hartree-Fock calculation.}
\label{table:t1}
\end{deluxetable}
\begin{deluxetable}{cccccccc}
\tablecolumns{8}
\tablewidth{0pt}
\tabletypesize{\footnotesize}
\tablecaption{P~{\small II} Branching Fractions for 3$s$3$p^3$ $^3P^{\rm o}$ Levels}
\tablehead{\colhead{Transition} & \colhead{ $J_l$} & \colhead{Present\tablenotemark{a}} &\colhead{ H\tablenotemark{b}}& \colhead{T\tablenotemark{c}}& \colhead{FTI\tablenotemark{d}}& \colhead{BB\tablenotemark{e}} & \colhead{LS coupling}
}  
\startdata
$^3P_J$ $-$  $^3P_1^{\rm o}$ & 0 & $0.36 \pm 0.03$ & 0.346 & 0.339 & 0.332 & $\ldots$ & 0.333  \\ 
& 1 & $0.27 \pm 0.03$ & 0.276 & 0.270 & 0.254 & $\ldots$ & 0.250 \\ 
& 2&$0.37 \pm 0.04$ & 0.378 & 0.391 & 0.414 &  $\ldots$ & $0.417$ \\ 
$^3P_J$ $-$  $^3P_2^{\rm o}$ & 1 & $0.22 \pm 0.02$ & 0.242 & 0.241 & 0.244 & 0.318& 0.250  \\ 
& 2 & $0.78 \pm 0.05$ & 0.758 & 0.759 & 0.756 & 0.682& 0.750  \\
\
\enddata
\tablenotetext{a}{This work -- beam-foil experiment.}
\tablenotetext{b}{\citet{hib88}.}
\tablenotetext{c}{\citet{tay03}.}
\tablenotetext{d}{\citet{fro06}.}
\tablenotetext{e}{\citet{ban11} $-$ semi-empirical method.}
\label{table:t2}
\end{deluxetable}

\begin{deluxetable}{cccccccccccccc}
\tablecolumns{14}
\tablewidth{0pt} 
\tabletypesize{\footnotesize}
\tablecaption{P~{\small II} Oscillator Strengths for the Multiplet 
3$s^2$3$p^2$ $^3P$ $-$ 3$s$3$p^3$ $^3P^{\rm o}$}
\tablehead{\colhead{$\lambda_{ul}$ (\AA)} & \colhead{$J_l$} & \colhead{$J_u$} & \multicolumn{10}{c}{$f$-value ($\times 10^{-3}$)} \\
\cline{4-14} \\
\colhead{} & \colhead{} &  \colhead{} & \colhead{Present} & \colhead{SL\tablenotemark{a}} & \colhead{CMB\tablenotemark{b}}&\colhead{BS\tablenotemark{c}}& \colhead{LKI\tablenotemark{d}}&\colhead{S\tablenotemark{e}} &  \colhead{H\tablenotemark{f}} &\colhead{BMF\tablenotemark{g}} & \colhead{T\tablenotemark{h}} &\colhead{FTI\tablenotemark{i}} & \colhead{M\tablenotemark{j}}
}
\startdata
1310.70 & 2 & 2 & $13.8 \pm 1.0$ & $\ldots$ & $\ldots$ & $\ldots$ & $\ldots$ & $\ldots$ & 8.4\tablenotemark{k} & $\ldots$ & 14.6\tablenotemark{k} & 15.1 & 8.4 \\
{       }& $\ldots$ & $\ldots$ & $\ldots$ & $\ldots$ & $\ldots$ & $\ldots$ & $\ldots$ & $\ldots$ & 8.0\tablenotemark{l} & $\ldots$ & 14.6\tablenotemark{l} & $\ldots$ & $\ldots$ \\
1309.87 & 2 & 1 & $4.1 \pm 0.5$ & $\ldots$ & $\ldots$ & $\ldots$ & $\ldots$ & $\ldots$ & 2.8\tablenotemark{k} & $\ldots$ & 4.8\tablenotemark{k} & 5.3 & 2.8 \\
{       } & $\ldots$ & $\ldots$ & $\ldots$ & $\ldots$ & $\ldots$ & $\ldots$ & $\ldots$ & $\ldots$ & 2.6\tablenotemark{l} & $\ldots$ & 4.8\tablenotemark{l} & $\ldots$ & $\ldots$ \\
1305.50 & 1 & 2 & $6.4 \pm 0.6$ & $\ldots$ & $\ldots$ & $\ldots$ & $\ldots$ & $\ldots$ & 4.5\tablenotemark{k} & $\ldots$ & 7.6\tablenotemark{k} & 8.1 & 4.5 \\
{       }& $\ldots$ & $\ldots$ & $\ldots$ & $\ldots$ & $\ldots$ & $\ldots$ & $\ldots$ & $\ldots$ & 4.5\tablenotemark{l} & $\ldots$ & 7.7\tablenotemark{l} & $\ldots$ & $\ldots$ \\
1304.68 & 1 & 1 & $4.9 \pm 0.6$ & $\ldots$ & $\ldots$ & $\ldots$ & $\ldots$ & $\ldots$ & 3.4\tablenotemark{k} & $\ldots$ & 5.5\tablenotemark{k} & 5.4 & 3.4 \\
{       } & $\ldots$ & $\ldots$ & $\ldots$ & $\ldots$ & $\ldots$ & $\ldots$ & $\ldots$ & $\ldots$ & 3.3\tablenotemark{l} & $\ldots$ & 5.6\tablenotemark{l} & $\ldots$ & $\ldots$ \\
1304.49 & 1 & 0 & $\ldots$ & $\ldots$ & $\ldots$ & $\ldots$ & $\ldots$ & $\ldots$ & 4.5\tablenotemark{k} & $\ldots$ & 7.0\tablenotemark{k} & 7.2 & 4.5 \\
{       } & $\ldots$ & $\ldots$ & $\ldots$ & $\ldots$ & $\ldots$ & $\ldots$ & $\ldots$ & $\ldots$ & 4.4\tablenotemark{l} & $\ldots$ & 7.0\tablenotemark{l} & $\ldots$ & $\ldots$ \\
1301.87 & 0 & 1 & $19.6 \pm 2.0$ & $\ldots$ & $\ldots$ & $\ldots$ & $\ldots$ & $\ldots$ & 12.7\tablenotemark{k} & $\ldots$ & 20.7\tablenotemark{k} & 21.0 & 12.7 \\
{       } & $\ldots$ & $\ldots$ & $\ldots$ & $\ldots$ & $\ldots$ & $\ldots$ & $\ldots$ & $\ldots$ & 12.6\tablenotemark{l} & $\ldots$ & 21.0\tablenotemark{l} & $\ldots$ & $\ldots$ \\
Multiplet & $\ldots$ & $\ldots$ & $\ldots$ & $40 \pm 5$ & $29 \pm 2$ & 38.2\tablenotemark{k} & $17 \pm 5$ & 48 & 11.8\tablenotemark{k} & 22.6\tablenotemark{k} & $\ldots$ & $\ldots$ & 11.8 \\
{       } & $\ldots$ & $\ldots$ & $\ldots$ & $\ldots$ & $\ldots$ & 43.6\tablenotemark{l} & $\ldots$ & $\ldots$ & 11.4\tablenotemark{l} & 23.0\tablenotemark{l} & $\ldots$ & $\ldots$ & $\ldots$ \\
\
\enddata
\tablenotetext{a}{\citet{sav66}.}
\tablenotetext{b}{\citet{cur71}.}
\tablenotetext{c}{\citet{bcs72} $-$ nonclosed many-electron theory.}
\tablenotetext{d}{\citet{liv75}.}
\tablenotetext{e}{\citet{smi78}.}
\tablenotetext{f}{\citet{hib88}.}
\tablenotetext{g}{\citet{bra93}.}
\tablenotetext{h}{\citet{tay03}.}
\tablenotetext{i}{\citet{fro06}.}
\tablenotetext{j}{\citet{mor03} $-$ compilation.}
\tablenotetext{k}{Based on length formalism.}
\tablenotetext{l}{Based on velocity formalism.}
\label{table:t3}
\end{deluxetable}
\clearpage

\begin{figure}[ht]
\centering
\includegraphics[width=0.7\textwidth]{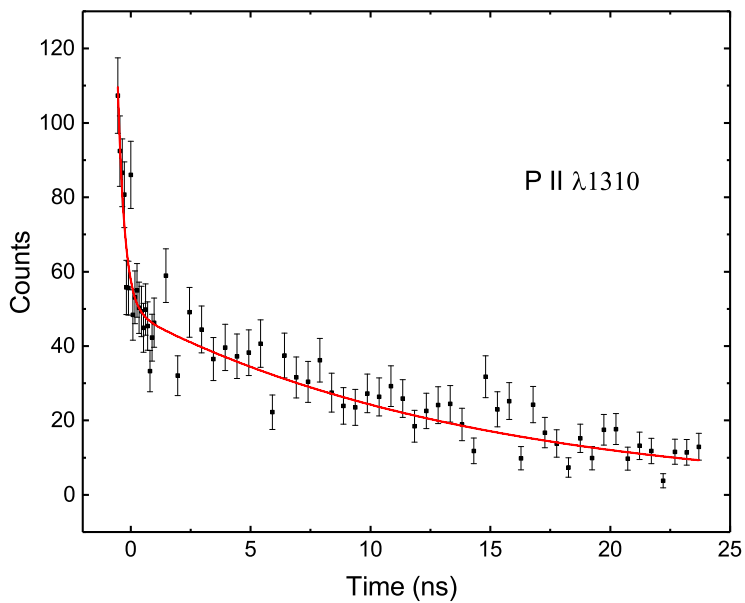}
\caption{The measured P~{\small II} decay curve for the line at 1310 \AA\ for a beam energy of 170 keV.  The post-foil beam velocity at this energy was 1.0068 mm ns$^{-1}$, thus establishing the time since excitation for a given foil position.  The foil was moved relative to the monochromator entrance slit in increments of 0.1 mm until it was displaced 1 mm; then the increments were increased to 0.5 mm. A two-exponential fit to the data is shown by the solid curve.}
\label{fig:f1}
\end{figure}

\begin{figure}[ht]
\centering
\includegraphics[width=0.7\textwidth]{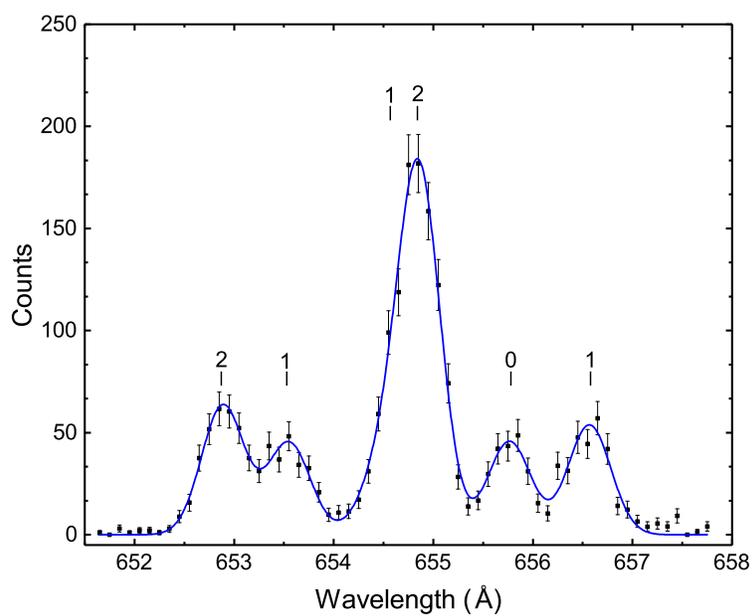}
\caption{A spectrum of the P~{\small IV} multiplet at 655 \AA\ for a beam energy of 205 keV. The total angular momentum quantum number for each upper fine structure level is indicated. Numbers above the spectrum refer to $J_u$ for different transitions. The best fit to the lines in the multiplet is shown by the solid curve.}
\label{fig:f2}
\end{figure}

\clearpage

\begin{figure}
\begin{center}
\includegraphics[width=0.7\textwidth]{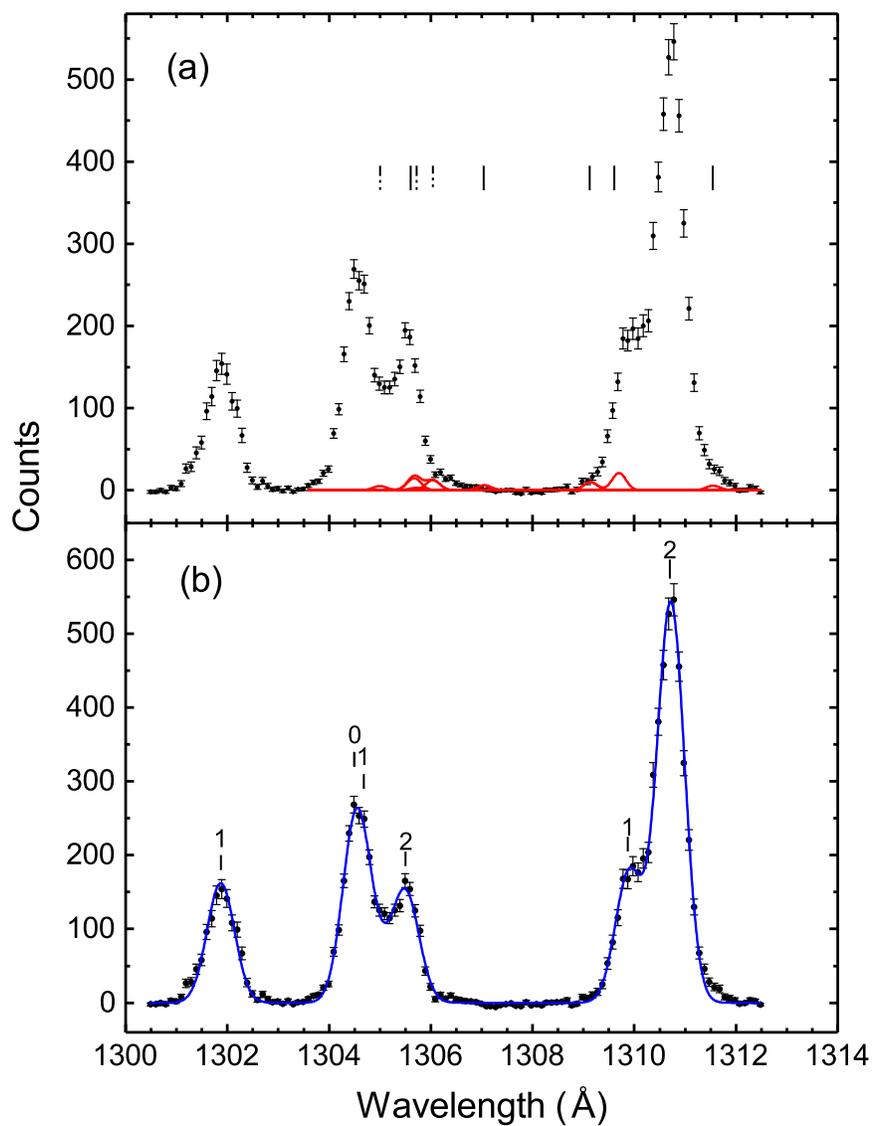}
\end{center}
\vspace{0.3in}
\caption{Panel (a) shows the uncorrected spectrum of the P~{\footnotesize II} multiplet at 1308 \AA\ with contaminating features. Dash-dotted tick marks indicate the position of P~{\small I} lines. Solid tick marks show the position of P~{\footnotesize IV} lines. Gaussian profiles reveal the strength of the contaminating features. Panel (b) represents to the corrected spectrum of the P~{\footnotesize II} multiplet at 1308 \AA. See Fig. 2 for further details.}
\label{fig:f3}

\end{figure}
%
\end{document}